\documentstyle[twocolumn,aps,prl,epsf]{revtex}
\begin{document}
\title{
Stripes Have Hair}

\author{R.S. Markiewicz}

\address{Physics Department and Barnett Institute, 
Northeastern U.,
Boston MA 02115}
\maketitle
\begin{abstract}
It is suggested that recent experiments provide evidence for modulations
{\it along} the charged stripes in the striped phase of the cuprates.
Furthermore, 1/8 doping is special because it is associated with a crossover:
for lower doping the magnetic stripe width changes; for higher, the hole-doped
stripes grow.  The crossover is reflected in the doping dependence of the spin 
gap.
\end{abstract}

\narrowtext

Recently, McQueeney, et al.\cite{EgP} reexplored the phonon dispersion in 
La$_{2-x}$Sr$_{x}$CuO$_4$ (LSCO), x=0.15.  They found a striking anomaly in the 
LO phonon branch associated with planar oxygen bond stretching: `a sharp
discontinuity (15 meV) at (0.25,0,0) at low temperatures.'  They interpreted 
this as evidence for zone folding due to a period doubling in the (1,0,0) 
direction.  Since static superlattice diffraction peaks are absent, they
suggest that this must represent a dynamic instability.  
\par
There is a great temptation to associate this with the striped phases found in 
the cuprates\cite{Tran}, particularly since the stripes are believed to be
dynamic at x=0.15\cite{Tran2}.  However, the stripes have the wrong periodicity:
at x=0.125, the stripes have a repeat distance of four unit cells, which would
correspond to period quadrupling, not doubling.  With enhanced doping the
periodicity changes, but no obvious period doubling is expected near $x=0.15$.
\par
In a photoemission study of optimally doped Bi$_2$Sr$_2$CaCu$_2$O$_8$, (Bi-2212)
Shen, et al.\cite{ZX} found a large transfer of spectral weight taking place
below the superconducting transition temperature -- which at optimal doping
coincides with the pseudogap onset temperature.  This spectral weight transfer
seems to be associated with a wave vector $Q^{\prime}=(\eta\pi ,0)$, with
$\eta\simeq 0.4-0.5$, and was also suggested to be associated with stripe 
formation.
\par
Strong fluctuation effects rapidly wash out interstripe correlations, but 
intrastripe correlations can be much more robust, particularly if these 
correlations help stabilize the striped phases. In fact, a number of models for 
the stripes involve charge density modulations along the hole-doped stripes 
which could generate a period doubling along (1,0,0).  Here, I compare these 
models.
\par
Hartree-Fock calculations\cite{HF} of the tJ model find that the holes condense 
onto domain walls between antiferromagnetically ordered domains, producing fully
occupied charge stripes -- one hole per Cu -- and hence no modulation along the 
stripe.  However, neutron diffraction\cite{Tran} finds a charge modulation of
periodicity four Cu atoms at $x=0.125$, which implies only 1/2 hole per cell.
Tranquada, et al.\cite{Tran} suggested a model for the charged stripes, based on
their experience with stripes in nickelates.  The hole-doped stripes are one 
cell wide, and have a hole on every other site.  A microscopic model for such a 
domain wall can be derived\cite{Zaa3} by incorporating a charge-density wave 
(CDW) instability along the stripes, treating them as one-dimensional metals.
\par
Such an arrangement would
have an extremely strong coupling to (1,0,0) period doubling.  (This has also
been suggested by Egami, et al.\cite{ELM}.)  Note 
that this requires a restriction on possible interstripe correlations.  If the
interstripe correlations were dominated by hole-hole Coulomb repulsion, one
might expect that adjacent stripes would align with the holes offset, to
enhance the average hole-hole separation.  However, the $(l,0,0)$
structure is sensitive only to the average structural modulation along the 
x-axis, so such an offset structure would not display any period doubling along
x.  On the other hand, if the stripes are stabilized by elastic forces, there is
no reason why holes on adjacent stripes could not line up.  Indeed, offset holes
on adjacent chains would cause a period doubling (from 4 cells to 8) along the
stripe repeat direction, which is not seen experimentally\cite{Tran,xray}.
\par
\begin{figure}
\setlength{\unitlength}{1.mm}
\begin{picture}(60,40)
\footnotesize
\put(10,15){\circle{2}}
\put(10,20){\circle{2}}
\put(10,25){\circle{2}}
\put(15,15){\circle*{2}}
\put(15,20){\circle{2}}
\put(15,25){\circle*{2}}
\put(20,15){\circle{2}}
\put(20,20){\circle{2}}
\put(20,25){\circle{2}}
\put(25,15){\circle{2}}
\put(25,20){\circle{2}}
\put(25,25){\circle{2}}
\put(30,15){\circle*{2}}
\put(30,20){\circle{2}}
\put(30,25){\circle*{2}}
\put(15,16){\line(0,1){3}}
\put(15,21){\line(0,1){3}}
\put(30,16){\line(0,1){3}}
\put(30,21){\line(0,1){3}}
\put(16,15){\line(1,0){3}}
\put(16,25){\line(1,0){3}}
\put(21,15){\line(1,0){3}}
\put(21,25){\line(1,0){3}}
\put(26,15){\line(1,0){3}}
\put(26,25){\line(1,0){3}}
\put(40,10){\circle*{2}}
\put(40,20){\circle{1}}
\put(40,30){\circle*{2}}
\put(60,10){\circle*{2}}
\put(60,20){\circle{1}}
\put(60,30){\circle*{2}}
\put(46.7,10){\circle{1}}
\put(46.7,20){\circle{1}}
\put(46.7,30){\circle{1}}
\put(53.3,10){\circle{1}}
\put(53.3,20){\circle{1}}
\put(53.3,30){\circle{1}}
\normalsize
\put(4,37){\makebox(0,0){\bf (a)}}
\put(39,37){\makebox(0,0){\bf (b)}}
\end{picture}
\caption{(a) (1,2) superlattice; box = unit cell. (b) Diffraction pattern,
showing superlattice spots.  Filled circles = original tetragonal lattice.  All 
superlattice spots have the same relative intensity.}
\label{fig:5}
\end{figure}
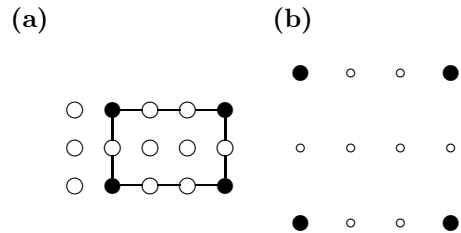

Figure~\ref{fig:5} illustrates the superlattice associated with these Tranquada 
stripes and the associated diffraction pattern, for a nearby commensurate doping
$x=1/6$.  While
there are superlattice spots at $(1/2,0)$, there are a number of other spots
which have not been detected.  The stripe phases are labelled $(N,M)$, where
$N$ is the width of the hole-doped stripes, and $M$ is the width of the
magnetic (undoped) stripes.  The striped phase of Fig.~\ref{fig:5}
corresponds to (1,2).  

\par
While the Tranquada model suggests a plausible origin for the observed period 
doubling, recent Monte Carlo renormalization group calculations of the tJ model 
by White and Scalapino (WS)\cite{WhiSc} fail to find Tranquada stripes. Instead,
they find wider hole-doped stripes with only 0.18 holes per Cu in the doped 
stripes (the remaining charge is in the magnetic stripes, 0.07 hole per Cu).  
For $x\le 1/8$, the hole-doped stripes are two cells wide, while the magnetic
stripes change width with doping.  In a domain-wall picture, these stripes are 
interpreted in terms of `bond-centered' rather than `site-centered' 
holes\cite{Zaa2}.  
\par
If there is a modulation of the hole density on alternate bonds along the 
stripes, then a picture entirely analogous to Fig.~\ref{fig:5} will produce a
similar period doubling.  However, it seems more natural that Coulomb
repulsion would cause the holes to alternate on adjacent rows -- this would
have the effect of keeping all holes on the same magnetic sublattice.  While
the WS calculations do not reveal any obvious charge modulation along the 
hole-doped stripes, there is evidence for such diagonal pairing\cite{diag} for a
single pair of holes.  A similar hole modulation can be stabilized by strong 
electron-phonon coupling\cite{RM3,YZGB}. In a recent slave boson 
calculation\cite{Pstr}, a particular phonon coupling was assumed: a 
charge-density wave (CDW) at $\vec Q=(\pi ,\pi )$ with accompanying oxygen 
breathing mode distortion as Peierls instability.  
\par
While this modulation produces a period-doubling, it does not contribute to the
(1/2,0,0) spot, for the reasons discussed above, when the hole-doped stripes are
two cells wide.  Whenever the hole-doped stripe is
an odd number of cells wide, there will be an excess of charge in every other 
row, and hence a period doubling along (1,0,0) -- just as for the Tranquada
stripe.  Note in particular that if the Peierls distortion is associated 
with the oxygen breathing mode there should be a particularly strong effect on
the oxygen bond stretching mode, as observed.

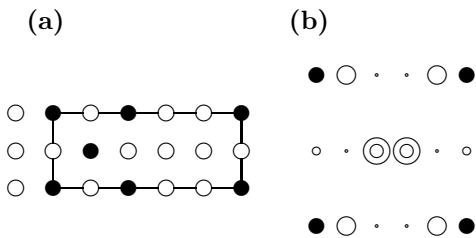
\begin{figure}
\setlength{\unitlength}{1.mm}
\begin{picture}(60,40)
\footnotesize
\put(00,15){\circle{2}}
\put(00,20){\circle{2}}
\put(00,25){\circle{2}}
\put(05,15){\circle*{2}}
\put(05,20){\circle{2}}
\put(05,25){\circle*{2}}
\put(10,15){\circle{2}}
\put(10,20){\circle*{2}}
\put(10,25){\circle{2}}
\put(15,15){\circle*{2}}
\put(15,20){\circle{2}}
\put(15,25){\circle*{2}}
\put(20,15){\circle{2}}
\put(20,20){\circle{2}}
\put(20,25){\circle{2}}
\put(25,15){\circle{2}}
\put(25,20){\circle{2}}
\put(25,25){\circle{2}}
\put(30,15){\circle*{2}}
\put(30,20){\circle{2}}
\put(30,25){\circle*{2}}
\put(05,16){\line(0,1){3}}
\put(05,21){\line(0,1){3}}
\put(30,16){\line(0,1){3}}
\put(30,21){\line(0,1){3}}
\put(06,15){\line(1,0){3}}
\put(06,25){\line(1,0){3}}
\put(11,15){\line(1,0){3}}
\put(11,25){\line(1,0){3}}
\put(16,15){\line(1,0){3}}
\put(16,25){\line(1,0){3}}
\put(21,15){\line(1,0){3}}
\put(21,25){\line(1,0){3}}
\put(26,15){\line(1,0){3}}
\put(26,25){\line(1,0){3}}
\put(40,10){\circle*{2}}
\put(40,20){\circle{1}}
\put(40,30){\circle*{2}}
\put(60,10){\circle*{2}}
\put(60,20){\circle{1}}
\put(60,30){\circle*{2}}
\put(44,10){\circle{2.6}}
\put(44,20){\circle{0.1}}
\put(44,30){\circle{2.6}}
\put(48,10){\circle{0.4}}
\put(48,20){\circle{3.4}}
\put(48,20){\circle{1.7}}
\put(48,30){\circle{0.4}}
\put(52,10){\circle{0.4}}
\put(52,20){\circle{3.4}}
\put(52,20){\circle{1.7}}
\put(52,30){\circle{0.4}}
\put(56,10){\circle{2.6}}
\put(56,20){\circle{0.1}}
\put(56,30){\circle{2.6}}
\normalsize
\put(4,37){\makebox(0,0){\bf (a)}}
\put(39,37){\makebox(0,0){\bf (b)}}
\end{picture}
\caption{(a) (3,2) superlattice; box = unit cell. (b) Superlattice spots.  
Filled circles = original tetragonal lattice.  For the superlattice spots, the 
relative intensity is proportional to the radius, except for the double circles,
for which the intensity must be doubled.}
\label{fig:3}
\end{figure}
\par
In order to interpret the experimental observations, I postulate a sort of 
percolation crossover of the striped phase at 1/8 doping.  Spin gap effects
cause the magnetic stripes to be preferentially an even number of Cu atoms
wide\cite{TTR} (this will be discussed further below).  Hence at x=0.125, the 
magnetic stripes have their narrowest width, two Cu's.  Additional hole doping 
can be accomodated for by making the hole-doped stripes wider, so that at x=0.15
there should be a certain number of 3-cell-wide hole-doped stripes, consistent 
with the present analysis.  However, for $x<0.125$, it will be the magnetic 
stripes which grow in width, the hole-doped stripes being fixed at their lowest 
width.  Thus, the (1,0,0) modulation should {\it disappear} for $x\le 0.125$.  
A study of the phonons in this low-doping regime can thus establish the nature 
of the modulation of the minimum charged stripes.
\par
Assuming LSCO to consist of magnetic stripes with 0.07 hole doping and charged
stripes with 0.18 hole, the material used in McQueeney, et al.\cite{EgP}'s
experiment, with $x=0.15$, would have a mixture of three and four cell-wide
hole stripes.  In Fig.~\ref{fig:3}a, I illustrate the nearby ($x=0.136$)
commensurate structure, for which all hole-doped stripes are three cells wide.
The resulting superlattice diffraction pattern reveals strong peaks near the
orthorhombic instability at (1/2,1/2), but also a significant peak near (1/2,0),
Fig.~\ref{fig:3}b.  Thus, since LSCO is orthorhombic, the present 
model produces only the superlattice spots needed to explain the experiments.

\begin{figure}
\setlength{\unitlength}{1.mm}
\begin{picture}(80,55)
\footnotesize
\put(00,45){\circle{2}}
\put(00,50){\circle{2}}
\put(00,55){\circle{2}}
\put(05,45){\circle*{2}}
\put(05,50){\circle{2}}
\put(05,55){\circle*{2}}
\put(10,45){\circle{2}}
\put(10,50){\circle*{2}}
\put(10,55){\circle{2}}
\put(15,45){\circle*{2}}
\put(15,50){\circle{2}}
\put(15,55){\circle*{2}}
\put(20,45){\circle{2}}
\put(20,50){\circle{2}}
\put(20,55){\circle{2}}
\put(25,45){\circle{2}}
\put(25,50){\circle{2}}
\put(25,55){\circle{2}}
\put(30,45){\circle*{2}}
\put(30,50){\circle{2}}
\put(30,55){\circle*{2}}
\put(35,45){\circle{2}}
\put(35,50){\circle*{2}}
\put(35,55){\circle{2}}
\put(40,45){\circle{2}}
\put(40,50){\circle{2}}
\put(40,55){\circle{2}}
\put(45,45){\circle{2}}
\put(45,50){\circle{2}}
\put(45,55){\circle{2}}
\put(50,45){\circle{2}}
\put(50,50){\circle*{2}}
\put(50,55){\circle{2}}
\put(55,45){\circle*{2}}
\put(55,50){\circle{2}}
\put(55,55){\circle*{2}}
\put(60,45){\circle{2}}
\put(60,50){\circle*{2}}
\put(60,55){\circle{2}}
\put(65,45){\circle*{2}}
\put(65,50){\circle{2}}
\put(65,55){\circle*{2}}
\put(70,45){\circle{2}}
\put(70,50){\circle{2}}
\put(70,55){\circle{2}}
\put(75,45){\circle{2}}
\put(75,50){\circle{2}}
\put(75,55){\circle{2}}
\put(80,45){\circle*{2}}
\put(80,50){\circle{2}}
\put(80,55){\circle*{2}}
\put(30,46){\line(0,1){3}}
\put(30,51){\line(0,1){3}}
\put(80,46){\line(0,1){3}}
\put(80,51){\line(0,1){3}}
\put(31,45){\line(1,0){3}}
\put(31,55){\line(1,0){3}}
\put(36,45){\line(1,0){3}}
\put(36,55){\line(1,0){3}}
\put(41,45){\line(1,0){3}}
\put(41,55){\line(1,0){3}}
\put(46,45){\line(1,0){3}}
\put(46,55){\line(1,0){3}}
\put(51,45){\line(1,0){3}}
\put(51,55){\line(1,0){3}}
\put(56,45){\line(1,0){3}}
\put(56,55){\line(1,0){3}}
\put(61,45){\line(1,0){3}}
\put(61,55){\line(1,0){3}}
\put(66,45){\line(1,0){3}}
\put(66,55){\line(1,0){3}}
\put(71,45){\line(1,0){3}}
\put(71,55){\line(1,0){3}}
\put(76,45){\line(1,0){3}}
\put(76,55){\line(1,0){3}}
\put(30,10){\circle*{2}}
\put(30,20){\circle{1.4}}
\put(30,30){\circle*{2}}
\put(50,10){\circle*{2}}
\put(50,20){\circle{1.4}}
\put(50,30){\circle*{2}}
\put(34,10){\circle{2.8}}
\put(34,20){\circle{0.5}}
\put(34,30){\circle{2.8}}
\put(38,10){\circle{0.2}}
\put(38,20){\circle{3.7}}
\put(38,20){\circle{1.9}}
\put(38,30){\circle{0.2}}
\put(42,10){\circle{0.2}}
\put(42,20){\circle{3.2}}
\put(42,20){\circle{1.6}}
\put(42,30){\circle{0.2}}
\put(46,10){\circle{3.2}}
\put(46,20){\circle{0.1}}
\put(46,30){\circle{3.2}}
\normalsize
\put(4,37){\makebox(0,0){\bf (a)}}
\put(20,12){\makebox(0,0){\bf (b)}}
\end{picture}
\caption{(a) 2-4 defect (boxed) replacing a 3-3 unit. (b) Superlattice spots for
a regular array with a 2-4 replacing every fourth 3-3.}  
\label{fig:4}
\end{figure}
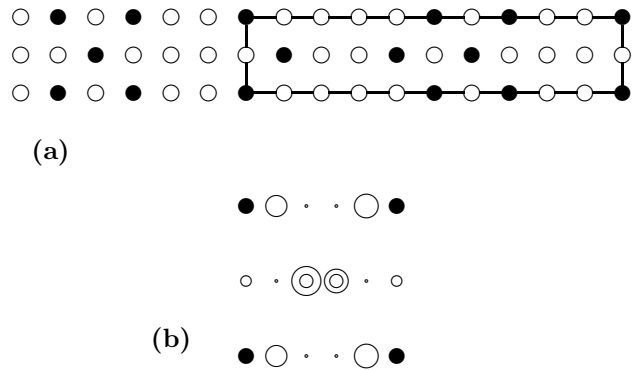

Stripe disorder will tend to consist of irregular
variations in the widths of the charged stripes, with the most likely pattern
being 3-3$\rightarrow$2-4, Fig.~\ref{fig:4}a.  As long as the 2-4 stripes
remain bound in pairs, they will not affect the phase of the 3-3 stripe pattern,
and hence will make relatively small changes in the superlattice pattern,
Fig.~\ref{fig:4}b.  Isolated 2 or 4 stripes on the other hand act as antiphase
boundaries, limiting the size of the 3-3-3... domains.

\par
\begin{figure}
\leavevmode
   \epsfxsize=0.33\textwidth\epsfbox{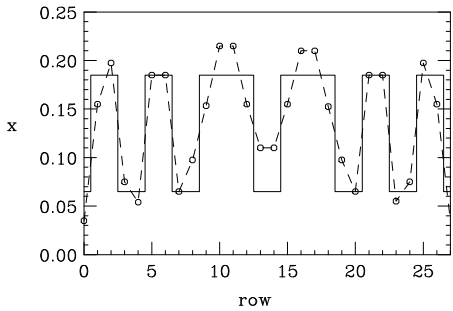}
\vskip0.5cm 
\caption{Charged stripes in the tJ model\protect\cite{WhiSc2}, interpreted as 
constant hole density domains of variable width.}
\label{fig:1}
\end{figure}

The above calculations require a form of {\it percolation crossover} to arise at
1/8 doping: for $x<(>)1/8$, the hole-doped (magnetic) stripes have a minimum 
width of two cells, while the magnetic (hole-doped) stripe width varies with 
doping.  Is there any direct evidence for such a crossover?  Recent calculations
by WS\cite{WhiSc2} find a crossover at 1/8 doping, which can be reinterpreted as
support for this picture.  Figure \ref{fig:1} replots the data of Fig. 4a in 
Ref.\cite{WhiSc2}; to minimize finite size effects, the circles represent the 
average hole density along a given row.  The solid lines illustrate an 
interpretation in terms of a phase separation picture.  There are just two 
preferred densities, low for spin stripes and high for hole stripes.  The 
crossover can be interpreted as a change in the hole-doped stripe width, with 
hole density remaining constant.  This explains why the crossover point is at 
1/8 filling: here the magnetic stripes have their minimum width, so adding extra
holes requires increasing the hole stripe width from its minimum value.  

If the phase separation model is valid, it should be possible to predict the
doping dependence of the stripe width.  If it is assumed that all the holes are 
associated with the hole-doped stripes, but in the low-doping limit some of the 
hole density leaks off the charged stripes into the adjacent magnetic layers, 
then the WS calculations find $x_0$=0.25 holes per Cu in the low doping limit. 
In this case, two-cell-wide hole-doped stripes should persist until a 
percolation crossover at $x=x_0/2=0.125$, the uniform 4-cell-wide stripes
should be commensurate at $x=2x_0/3$=0.167, and the stripe phase should 
terminate at $x=x_0=0.25$; WS find these phase boundaries at $x$ = 0.125, 0.17,
and 0.30, respectively.  The agreement is quite good, but there is evidence
that the hole-doped stripes might have a weakly doping-dependent hole density. 
Note that it is not clear why there is a preferred hole density in the tJ model.

\par
\begin{figure}
\leavevmode
   \epsfxsize=0.33\textwidth\epsfbox{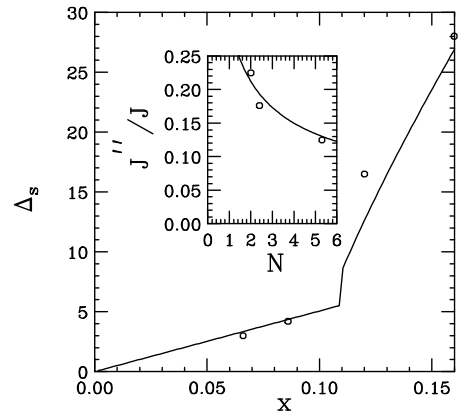}
\vskip0.5cm 
\caption{Spin gap $\Delta_s$ vs. doping for YBCO.  Circles = data of 
Ref.~\protect\cite{R-M}; line = theory, assuming solid line from inset. 
Inset: interladder exchange vs. hole-doped stripe width.}
\label{fig:7}
\end{figure}

Additional, experimental evidence for the crossover comes from the doping
dependence of the spin gap in YBa$_2$Cu$_3$O$_{7-\delta}$ (YBCO)\cite{R-M}.
A spin gap arises naturally in a spin ladder with an even number of legs, and
it has been suggested that such a gap preferentially stabilizes even-width
magnetic stripes\cite{TTR}.  The observed spin gap in YBCO can be interpreted
simply in terms of coupled magnetic ladders, Fig.~\ref{fig:7}.  Densities were
estimated assuming a parabolic $T_c(x)$ relationship\cite{Tal}, 
\begin{equation}
{T_c\over T_{c,max}}=1-({x-0.16\over 0.11})^2,
\label{eq:1c}
\end{equation}
corrected for a suppressed $T_c$ near the 60K plateau (i.e., assuming that the 
plateau is a form of 1/8 effect\cite{Tal2}).  Below the crossover, taken as 
$~0.11$, the 
magnetic stripe (ladder) width decreases smoothly with doping, while the 
interladder coupling is approximately constant, since the hole-doped stripe has 
fixed width.  Theoretically, the spin
gap is found to be (approximately) inversely proportional to the ladder
width\cite{RiD}, so in this regime the spin gap scales linearly with doping,
$\Delta_s=\beta J/M$, where $J=130meV$ is the exchange constant, $M$ the ladder
width, and $\beta$ a correction for interladder coupling, $\beta\simeq (1-
4J^{\prime\prime}/J)$, with $J^{\prime\prime}$ the exchange coupling between 
adjacent ladders\cite{Ric2}.  The solid line in Fig.~\ref{fig:7} corresponds to
$J^{\prime\prime}=0.225J$.
\par
Above the crossover, $x>x_0/2=0.11$, $M$ is fixed at 2 while $\beta$ increases 
with doping, since $J^{\prime\prime}$ decreases as the hole-doped stripes widen.
Since the Cu in the hole-doped stripes can be magnetized, the falloff should be
relatively slow.  Details are model sensitive, but qualitatively the observed
behavior is readily reproduced.  The curve in Fig.~\ref{fig:7} follows from
assuming a falloff $J^{\prime\prime}\sim N^{-1/2}$, where $N$ is the hole-doped
stripe width, inset to Fig.~\ref{fig:7}.  (It should be noted that the falloff
is sensitive to the hole-density $x_0$, here taken as 0.22.)
\par
In conclusion, recent experiments provide evidence that the hole-doped charge
stripes have `hair': an inhomogeneous structure along the chain which couples
strongly to phonons.  Furthermore, 1/8 doping is associated with a crossover
phenomenon, with the hole-doped stripes increasing in width at higher dopings.

I acknowledge a stimulating conversation with B.C. Giessen, L. Sakharov, M.
Sakharova, and C. Kusko.  Publication 747 of the Barnett Institute.

\end{document}